\documentclass[fleqn,usenatbib]{mnras}

\usepackage{newtxtext,newtxmath}
\usepackage{mathptmx}

\usepackage[T1]{fontenc}
\usepackage[switch]{lineno} 

\DeclareRobustCommand{\VAN}[3]{#2}
\let\VANthebibliography\thebibliography
\def\thebibliography{\DeclareRobustCommand{\VAN}[3]{##3}\VANthebibliography}

\usepackage{graphicx}	
\usepackage{amsmath}
\usepackage{tikz}

\usepackage{placeins}
\usepackage[english]{babel}
\usepackage{lmodern}
\usepackage{wasysym}  

\tikzstyle{process} = [rectangle, minimum width=4cm, minimum height=1cm, text centered, draw=black, fill=white!30]
\tikzstyle{decision} = [diamond, aspect=2, minimum width=3cm, minimum height=1cm, text centered, draw=black, fill=green!30]
\tikzstyle{arrow} = [thick,->,>=stealth]

\let\longtable*\relax

\title [The 2012 Venus transit]{Venus as an exoplanet analog: extended UV transit signatures and coronal occultations} 
\author[T. Pyne et al.]{
Tisyagupta Pyne,
Belur Ravindra,
Ravinder K. Banyal\thanks{E-mail: banyal@iiap.res.in}
\\
Indian Institute of Astrophysics, Koramangala 2nd Block, Bangalore 560034, India
}

\date{Accepted 2025 December 05. Received 2025 December 05; in original form 2025 October 17}

\pubyear{2025}

\begin{document}
\label{firstpage}
\pagerange{\pageref{firstpage}--\pageref{lastpage}}
\maketitle

\begin{abstract}
Stellar activity manifests differently across wavelengths, causing flux variability that can obscure planetary transits. While transit observations are typically performed in the visible and infrared bands, where stellar flux is relatively stable, short-wavelength regimes exhibit high variability, complicating reliable detections. Here, we analyze the 2012 transit of Venus as an exoplanet analog using multiwavelength observations taken by the Solar Dynamics Observatory (SDO) in five channels: 6173~\AA\ (continuum), 1700~\AA\ (broadband), and three extreme-UV (EUV) narrowbands at 304~\AA, 171~\AA, and 94~\AA. We find that the disk-integrated transit signal is clearly detectable in the 6173~\AA\ band, whereas strong solar activity-induced fluctuations obscure the transit in the EUV channels. Notably, the 1700~\AA\ UV transit is noisier but significantly longer ($\approx 9.2$~hrs) than the visible-band transit ($\approx 6.7$~hrs), because Venus began occulting the extended coronal features before ingress onto the visible disk. This observation highlights the potential of UV transits to probe the spatial extent of stellar coronae in exoplanetary systems. Numerical simulations further suggest that limb-brightened stars in quiescence phase may exhibit distinctive UV/EUV transit signatures, opening new possibilities for exoplanet detection and characterization in these spectral regimes. 

\end{abstract}


\section{Introduction}\label{sec1}
One of the earliest reports of a Venus transit was provided by Horrocks\footnote{\href{https://eclipse.gsfc.nasa.gov/OH/transit12.html}{eclipse.gsfc.nasa.gov/OH/transit12.html}} on December 4, 1639 after modifying the calculations put forward by Johannes Kepler in his ``Rudolphine Tables'' in 1631 \citep{Posch_2004,Chapman_2005,Powell2019}.
About 121.5 years later, Lomonosov observed the transit of Venus (ToV) in 1761 and suggested the presence of an atmosphere on Venus for the first time \citep{mAROV_2005}. Eight years later, in 1769, James Cook led an expedition to Tahiti to observe this event. They attempted to determine the solar parallax and find the size of the solar system and the distance between Earth and the Sun \citep{Orchiston_2005,Orchiston_2017}. The next pair of transits in 1874 and 1882 were again used to further investigate the parallax of the Sun and obtain more accurate values. By the end of the 20$^\textrm{th}$ century, advances in radar had enabled precise determinations of the astronomical unit (AU), diminishing the scientific returns from the upcoming transits in 2004 and 2012 \citep{Dick_1995}. 

However, in stellar astronomy, the discovery of the first exoplanet by \cite{Mayor_1995} using the radial velocity method, and the detection of the first transiting exoplanet by \cite{Cha_2000}, had already opened new prospects in planetary science.  In the subsequent years, transit photometry emerged as a promising new technique for exoplanet detection. This method analyzed the variation in stellar flux during a planetary transit and measured its dip to estimate the planetary radius relative to the star \citep{Charbonneau_2000}. Today, the transit method accounts for the detection of $\sim$75\% of all known exoplanets \citep{Christiansen_2025arXiv}. 

The ToV repeats in an interval of 8, 105.5, 8 and 121.5 years. After the succession of 2004 and 2012, the next pair of transits\footnote{\href{https://eclipse.gsfc.nasa.gov/transit/transit.html}{eclipse.gsfc.nasa.gov/transit/transit.html}} will be observed in December 11, 2117 and December 08, 2125 \citep{Garcia_2012}. In the 21$^\textrm{st}$ century, the ToV across the Sun has served as a key astronomical event for investigating the radius of the Sun \citep{Emilio_2015}, studying the planetary atmosphere of Venus \citep{Reale_2015}, exploring its transmission spectra, measuring the scattered light in solar telescopes and more \citep [see e.g.,][]{Ambastha_2006,Hedelt_2011,Garcia_2012,Sigismondi_2012,Prabhu_2013,Hauchecorne_2014}. 

Several space missions have also been sent to Venus, observing it up close via orbital, flyby and in situ missions \citep{Svedhem_2007b,Svedhem_2007a,Ignatiev_2009,Akatsuki_2011,Rosenblatt_2012,Akatsuki_2014,Reale_2015,Kane_2021,Venus_2022}. 
This makes Venus a well-studied planet in the solar system that can aid our understanding of exoplanet science. Accurate estimates of Venus's mass, radius, atmospheric composition, and other properties can serve as a valuable benchmark for the understanding and characterization of newly discovered exoplanets. \citep{Martin_Livio_2015,Horner_2020}. Furthermore, due to the detailed exploration of Venus and its similarity to Earth in terms of mass and radius, it also serves as an anchor point for studying exoplanetary habitability \citep{Venus_2022,Kane_2024,Pyne}. 

Although the Venus transits in 2004 and 2012 have motivated several studies, the photometric variation of the \textit{Sun as a star} during these events remains less explored. Such an investigation requires continuous observations of the full solar disk from a vantage point aligned with the Sun-Venus system. Any misalignment of the telescope, failure to capture the entire solar disk, or interruptions in observation can significantly hinder the analysis of the photometric light curve during the Venus transit.

The transit of June 8, 2004 was observed by telescope facilities both on ground and in space. Ground observatories such as the Global Oscillations Network Group \citep[GONG;][]{Harvey_1996}\footnote{This event was recorded from mainly 3 GONG Observatories: Learmonth, Teide (Instituto de Astrofisica de Canarias) and Udaipur. However, presently the intensity data are available for Learmonth and Teide.} have jointly recorded this event \citep{Ambastha_2006}. GONG is a network of 6 telescopes -- the Big Bear Solar Observatory, High Altitude Observatory, Learmonth Solar Observatory, Udaipur Solar Observatory, Instituto de Astrofisica de Canarias, and Cerro Tololo Interamerican Observatory. However, no single telescope was able to observe the entire event due to differences in local sunrise and sunset times. Additionally, photometrically sensitive events like planetary transits are susceptible to adverse weather, variable air mass, and other environmental conditions, all of which can introduce significant photometric errors. As a result, during the 2012 transit, the GONG telescopes, distributed across various geographic locations, were only able to capture partial segments of the event.

Space-based telescopes, by contrast, are free from these limitations and are therefore more reliable for sensitive photometric observations. They can continuously monitor the solar disk without atmospheric interference or observational constraints imposed by Earth's rotation causing day-night gap. Notably, the Solar and Heliospheric Observatory (SOHO) in space was not aligned with the Sun-Venus system and therefore did not capture the Transit of Venus in 2004 and 2012. While telescopes such as the Transition Region and Coronal Explorer (TRACE), the Solar X-ray Imager (SXI) onboard the Geostationary Operational Environmental Satellite (GOES) \citep{Ambastha_2006}, and Hinode \citep{Young_2022} were able to observe the Sun-Venus system, they lacked a full-disk view of the Sun throughout the entire transit period. To our knowledge, the Solar Dynamics Observatory \citep[SDO;][]{Pesnell_2012} was the only space-borne instrument to successfully record the full solar disk throughout the entire 2012 Venus transit.

In this work, we investigate the photometric variations of the Sun as a star during the Venus transit of June 5-6, 2012. The SDO  continuously monitored the full solar disk from its geosynchronous orbit at an altitude of approximately 36,000~km, providing a unique opportunity to study a solar system planetary transit light curve across multiple wavelengths. Such observations are important for inferring atmospheric scale heights \citep[including that of Venus; see][]{Reale_2015} and have been used to probe the transmission spectra of exoplanets to investigate their atmospheric properties \citep[e.g.,][]{Ehrenreich_2012,Delrez_2018}. 

Using SDO's multiwavelength data, we examine whether transit light curves can be distinctly detected at different wavelengths. Since each wavelength probes a different layer of the solar atmosphere, characterized by varying intensities and magnetic activity, this study allows us to assess the manifestation of planetary transits across stellar disk with spatial and spectral inhomogeneity. It is important to note that the 2012 transit of Venus occurred near the peak of Solar Cycle 24, which means the Sun was in a relatively active phase of its 11-year cycle.

SDO observed the 2012 Venus transit using the Atmospheric Imaging Assembly \citep[AIA;][]{Lemen_2012} and the Helioseismic and Magnetic Imager \citep[HMI;][]{Schou_2012, Scherrer_2012} instruments onboard. We detrended and analyzed light curves constructed from disk-integrated images in different wavelength channels.

From here on, the paper is organized as follows. Section~\ref{sec:sdo_obs} describes the SDO observations and the multiwavelength data used in this study. The construction of disk-integrated light curves and the procedure for extracting the transit signal are detailed in Section~\ref{sec:lc_ext}. Our results and their interpretation are presented in Section~\ref{sec:result}. In Section~\ref{sec:tran_sim}, we extend the analysis through numerical simulations of transits across the solar disk under different activity levels to assess the effect of stellar variability on transit detectability. Finally, Section~\ref{sec:con}  summarizes the key findings and outlines their broader implications for exoplanet transit studies.

\begin{figure*}
  \centering
  \includegraphics[width=0.5\textwidth]{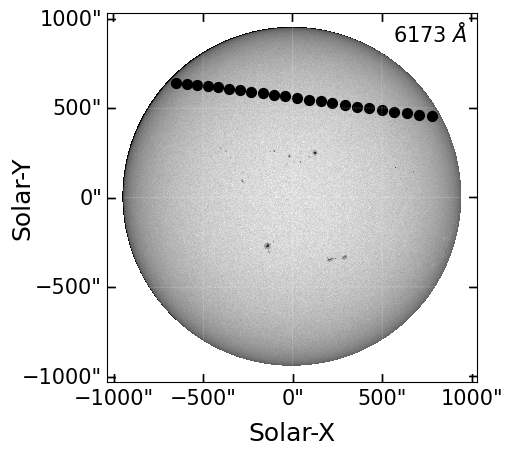}\hfill
  \includegraphics[width=0.5\textwidth]{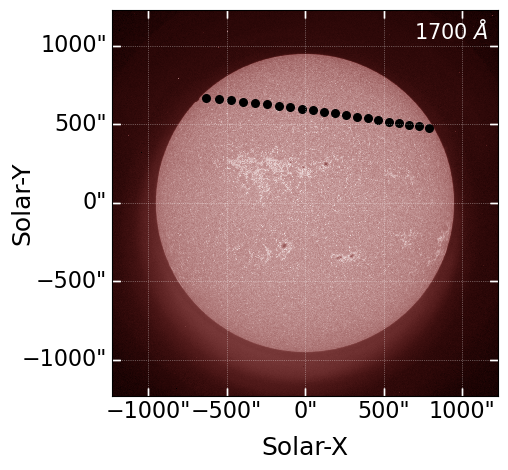}
  \vspace{0.6em}
  \includegraphics[width=0.33\textwidth]{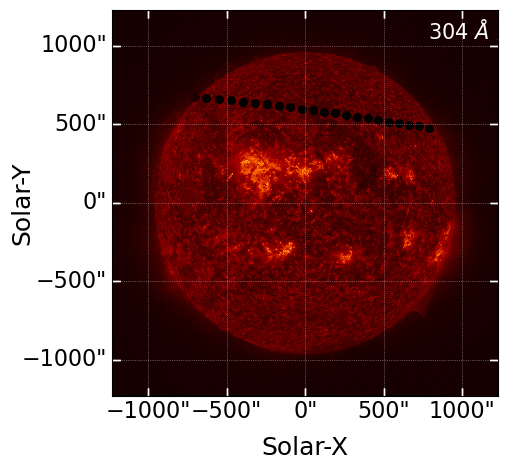}\hfill
  \includegraphics[width=0.33\textwidth]{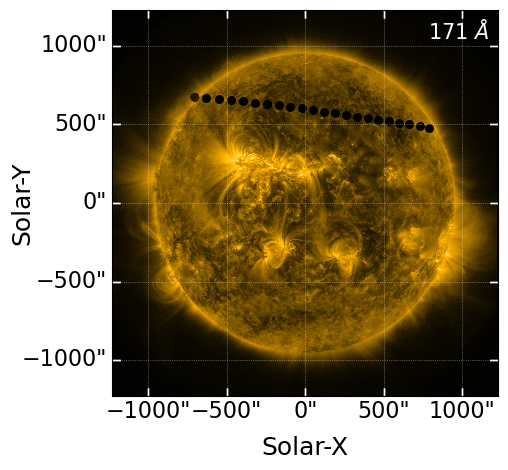}\hfill
  \includegraphics[width=0.33\textwidth]{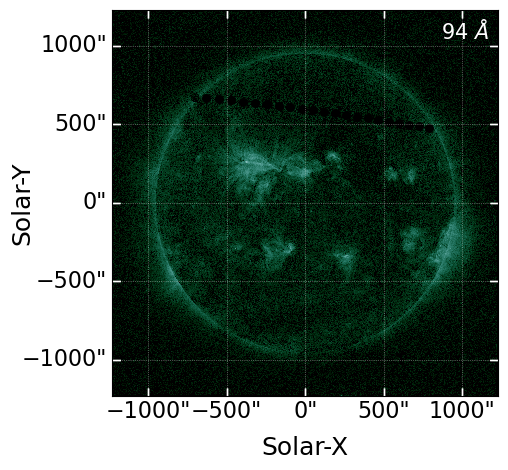}

  \caption{The transit of Venus across the sun on 5~June 2012 observed in 5~SDO channels. The black dots represent the path of Venus and do not depict the actual planet size or the cadence at which it was observed. Top row shows the images obtained in 6173\AA~ and 1700\AA~ channels; bottom row shows the images obtained in 304\AA, 171\AA, and 94\AA~ channels respectively.}
  \label{fig:transit-all}
\end{figure*}

\section{SDO observations  and data }\label{sec:sdo_obs}
The ToV, which occurred between 5$^\textrm{th}$ June 22:09 UTC and 6$^\textrm{th}$ June 04:49 UTC, 2012 was observed by the SDO from its circular geosynchronous orbit, $\sim$36000 km above Earth. The orbit is inclined by 28$^\circ$ about the longitude of the SDO-dedicated ground station in New Mexico. The SDO/HMI and AIA detectors have $4096 \times 4096$ pixels, with each pixel corresponding to $\sim0.5^{\prime\prime}$ on the sky for HMI and $\sim0.6^{\prime\prime}$ for AIA. The HMI observes the Sun at 6173~\AA~at a cadence of 45~s. The AIA observes the Sun at one whitelight (4500~\AA) channel at a cadence of 1~hour, two UV (1700~\AA, 1600~\AA) channels at a 24~s cadence and seven EUV (335~\AA, 304~\AA, 211~\AA, 193~\AA, 171~\AA, 131~\AA, 94~\AA) channels at a 12~s cadence. In total, the SDO primarily observes the full solar disk through these 11 channels. We use 5 of these channels onboard the SDO -- specifically: 6173~\AA\ (HMI Level 1.5)\footnote{see \cite{Scherrer_2012,Couvidat_2016} for further details on SDO/HMI data acquisition.}, 1700~\AA, 304~\AA, 171~\AA, and 94~\AA~(AIA Level 1.0)\footnote{Raw telemetry data (AIA Level 0) are converted to Level 1.0 after bad-pixel removal, despiking (for EUV channels only) and flat-fielding \citep{Lemen_2012}.} to study the ToV. 

Figure~\ref{fig:transit-all} shows the path of Venus across the solar disk, from left to right, as observed in five channels during the transit. The top panel indicates the 6173~\AA\ and 1700~\AA\ (left to right) whereas the bottom panel includes the 304~\AA, 171~\AA\ and 94~\AA~(left to right). 

\begin{figure*}
    \centering
    \includegraphics[width=\linewidth]{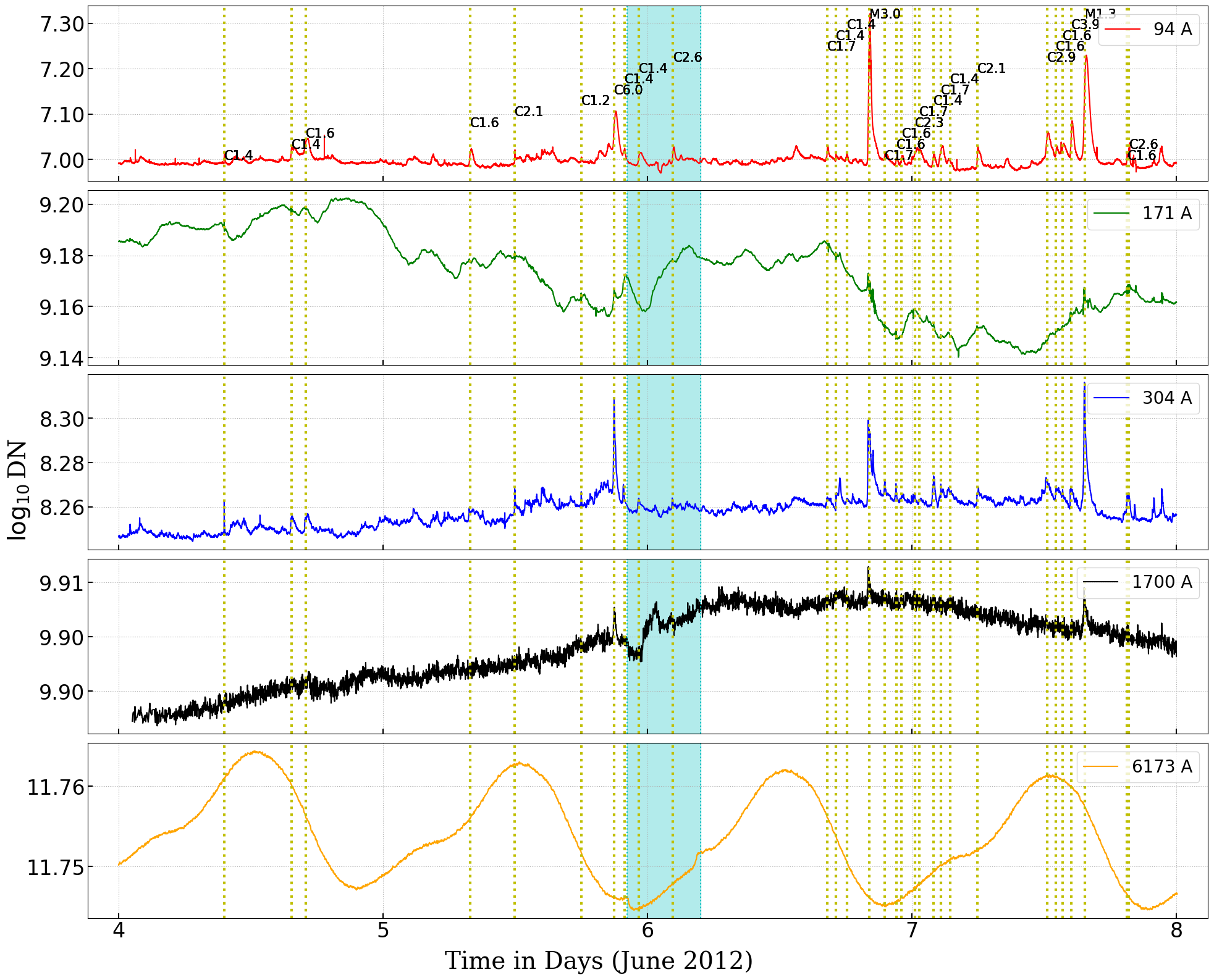}
    \caption{Disk-integrated light curves in five wavelengths bands from SDO's time series images taken between 4-7 June, 2012. A vertical cyan box marks the Venus transit, starting at 22:10~UTC on 5~June and ending at 04:50~UTC on 6~June 2012. The yellow dotted lines indicate the time when solar flares occurred during this 96~hours observation window. The strength of each flare is labeled to the right of the respective dotted lines in the top panel. Panels from \textit{top-to-bottom} corresponds to 94~\AA, 171~\AA, 304~\AA, 1700~\AA, 6173~\AA~ channel, respectively.}
    \label{fig:lc5}
\end{figure*}

We analyze full-disk solar images and constructed disk-integrated light curves in five wavelength bands from 00:00 UTC on 4 June to 00:00 UTC on 8 June 2012, covering $\sim$46 hours before the transit, $\sim$7 hours during the transit, and $\sim$43 hours after the transit. This extended timeline was chosen to capture  instrument systematics affecting the  transit light curve. Some AIA images, particularly those taken during the transit, were corrupted and had to be removed, which resulted in slightly uneven cadence (1–2 minutes). However,  sampling was still adequate for our analysis, so excluding these frames did not affect the results.\footnote{Note that all images of the 1600~\AA~ channel (AIA, Level 1.0) are corrupted within the duration of the transit.} In total, we used 26,300 of 28,300 images from the three EUV AIA channels, about 12,000 of 13,000 images from the UV channel, and 7,673 images from the HMI channel. 

The science-ready HMI Level~1.5 data include full-disk continuum images of the Sun, with all pixels outside the solar disk set to \texttt{NaN}. Each file header provides the SDO--Sun ephemeris and the time of observation. AIA images are designated as Level 1.0 data. We use \texttt{aiapy} \citep{Barnes2020} to promote the AIA Level 1.0 data to Level 1.5 by performing pointing corrections, degradation correction and registration. Further, images with \texttt{NaN} values were discarded and our sample was restricted to images with exposure time (\texttt{EXPTIME}) $\approx$~1~s for 1700~\AA\ and nearly 3~s for EUV wavelengths. 

\section{Disk-integrated light curves \& Transit extraction}\label{sec:lc_ext}
We integrate the data number (DN) over the full solar disk at a cadence of 45~s for HMI,  12~s for EUV wavelengths and 24~s for the UV wavelengths. The resulting value of each integrated image becomes a data point on the light curve. In Figure~\ref{fig:lc5} we show the raw light curves, obtained by integrating the solar disk intensity (expressed in DN), for the five SDO channels. The top 3 panels show the light curves from the data of 3~EUV AIA channels whereas the 4th panel shows the light curve obtained from 1700~\AA~UV channel. The bottom panel of Figure~\ref{fig:lc5} shows the light curve from the whitelight data at 6173~\AA. Furthermore, we obtained data from the Space Weather Prediction Center \footnote{\href{https://cesar.kso.ac.at/database/swpc_flares_query.php}{https://cesar.kso.ac.at/database/swpc\_flares\_query.php}}  of the National Oceanic and Atmospheric Administration (NOAA) to include the solar flares reported during the transit. Flare-related features in Figure~\ref{fig:lc5} are clearly absent in visible light curves but manifest distinctly in the UV and EUV regimes, highlighting the wavelength-dependent nature of flare emissions.

We note that instrumental systematics dominate the transit light curve. In the HMI 6173~\AA\  band we clearly identify two distinct trends: a sinusoidal modulation with a $\sim$24-hour period \citep[see][]{Couvidat_2016}, and a slow drift shifting the light curve downwards. The 24-hour component is likely related to spacecraft/diurnal orbital effects, while the slowly varying part may arise from thermal drift, sensitivity changes, or evolving background/pointing (its origin remains uncertain). 

\begin{figure*}
    \centering
    \includegraphics[width=\linewidth]{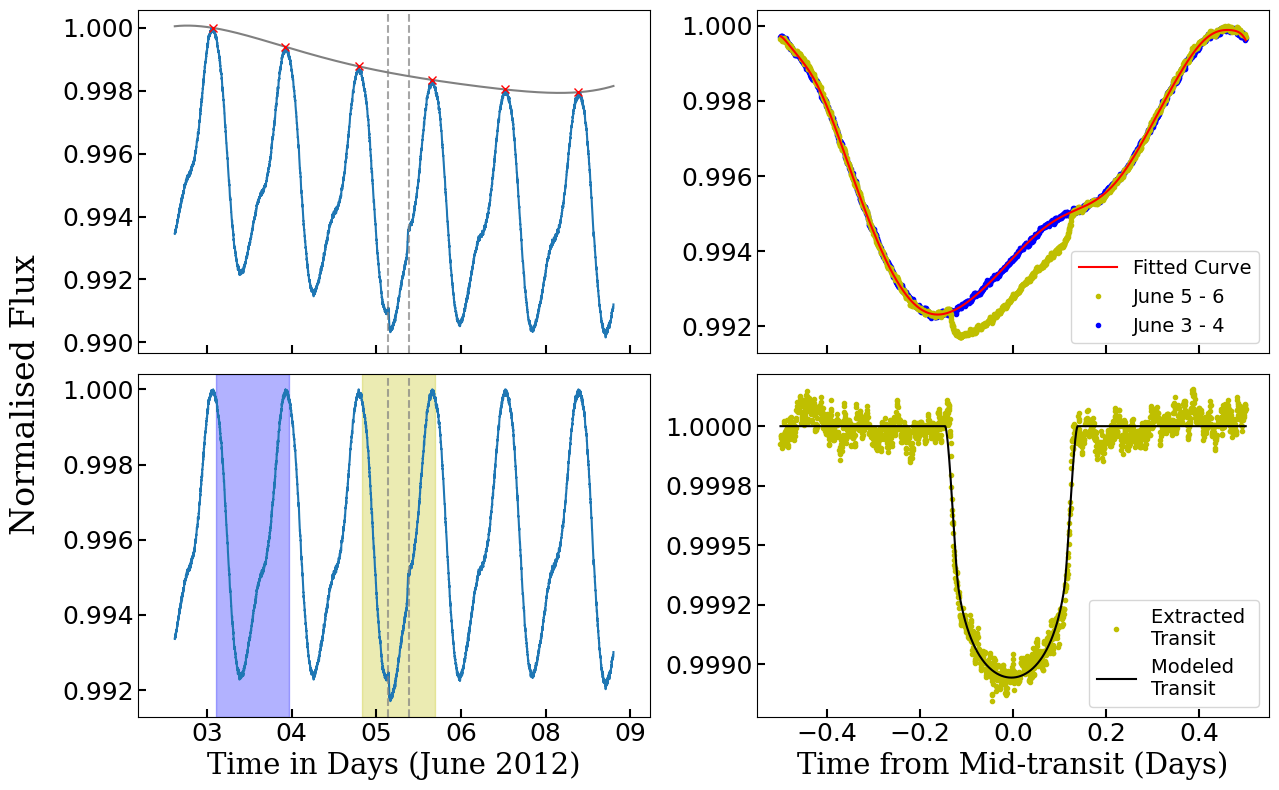}
    \caption{HMI 6173~\AA\ disk-integrated light curve and detrending steps. Top left: Raw disk-integrated light curve showing strong instrumental artifacts consisting of a 24-h periodic modulation and a slow downward drift. The gray curve is a low-order polynomial fit to peaks (red crosses mark) which models the slow drift.  Transit start and end times are indicated by vertical gray dashed lines. Bottom left: The drift-corrected light curve. A 24-h, transit-free reference segment (3-4 June, blue band) was chosen to model the periodic artifacts. A yellow band (5-6 June) denotes a 24 hours transit segment. Top right: Transit segment folded onto transit-free reference segment. A red curve is polynomial fit to transit-free reference segment. Bottom right: Extracted transit light curve from the polynomial fit and the transit segment. The black line indicates the best fit transit model obtained from python \texttt{batman-package}.}
    \label{fig:detrends}
\end{figure*}

To correct these systematics we proceed in two steps. First, we model the envelope of the peaks representing downward drift  with a third-order polynomial\footnote{The residuals continue to decrease with increasing polynomial order, but the improvement beyond the 16th order was marginal.} as shown in top-left panel of Figure~\ref{fig:detrends}, and divide the light curve by this fit to remove the slow trend. The resulting drift-free light curve is shown in bottom-left panel of Figure~\ref{fig:detrends}. Next, we fit a sixteen-order polynomial to a transit-free 24-hour reference segment (2012-06-03 13:29--2012-06-04 13:29 UTC; blue region, bottom-left panel) to model the 24-hour segment of periodic waveform, while masking the 24-hour transit epoch (2012-06-06 01:29 UTC; 2012-06-05 13:29--2012-06-06 13:29 UTC; yellow region, left panel of Figure~\ref{fig:detrends}). The masking was done to avoid biasing the transit depth. The transit signal was recovered by subtracting the polynomial fit to reference segment from the transit segment as shown in the right panel of Figure~\ref{fig:detrends}.  
Further, we fitted the detrended light curve with a standard transit model using the \texttt{batman-package} \citep{Mandel_2002,batman_2015} and used the resulting planet-to-star radius ratio to estimate the apparent angular radius of Venus relative to the Sun.

Likewise, we correct the 1700~\AA\ light curve using the same approach. After detrending the full four-day light curve with a fifteenth-order polynomial, we extract the transit segment from 2012-06-05 13:29--2012-06-06 13:29 UTC, which still shows a weak quadratic trend that we remove using a second-order polynomial before fitting the transit model.

Since only a single transit is available, we fit the parameters that are directly constrained by the data. The free parameters of the fit were: planet-to-star radius ratio, mid-transit time and constant flux offset. The orbital inclination was also allowed to vary but remained close to 90$^\circ$. All other parameters were fixed to the well-known values for Venus, including its orbital period (224.7 days), semi-major axis (0.723 AU), eccentricity (0), argument of periastron (131.5$^\circ$), and quadratic limb-darkening coefficients (0.3, 0.5). These parameters cannot be constrained from a single transit and fixing them avoids degeneracies.

\section{Results \& Discussions}\label{sec:result} 
\begin{figure}
    \centering
    \includegraphics[width=0.75\linewidth]{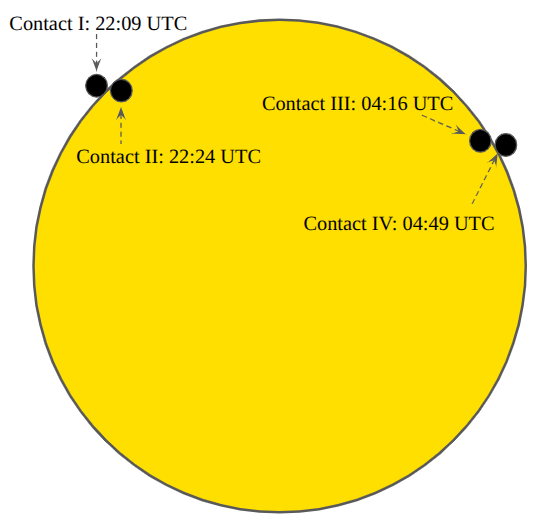}
    \caption{The 4 contact points during the ToV and their timings between June 5 and June 6, 2012. Note that the figure represents the definitions of the contact points and not the path of the transit.}
    \label{fig:contacts}
\end{figure}

We study the light curve of the Sun taken for 4 consecutive days -- 4$^\textrm{th}$ - 7$^\textrm{th}$ June 2012. Within this span, Venus enters AIA's field of view (FoV) around June 5, 20:30 UTC and ingresses on the solar disk around June 5, 22:09 UTC (Contact~I, see Figure~\ref{fig:contacts}). In SDO's FoV, it makes Contact II around 22:24 UTC and Contact III around 04:16 UTC. It egresses from the solar disk around June 6, 04:49 UTC (Contact IV) and exits AIA's FoV around June 6, 05:44 UTC.  

Unlike AIA, the HMI Level 1.5 data does not have Contact~I and Contact~IV information because all points outside the solar disk are set to \texttt{NaN}; therefore, only Contact II and Contact III are visible. While Venus is not prominent outside the disk in the UV images, off-limb solar features such as coronal loops -- visible in the EUV wavelengths -- cause Venus to appear outside the disk well before Contact I and after Contact IV. 
In this study, we primarily focus on the transit signatures in different wavelength bands. 
However, due to the high variability of the Sun in the EUV wavelengths, the ToV cannot be clearly extracted, even though Venus is visible off-limb in AIA's FoV. The following subsections discuss the light curves extracted from 6173~\AA\ and 1700~\AA\ channels in further detail. 

\subsection{ToV as an exoplanet analog}
From visual inspection of the 6173~\AA\ light curve, the transit appears as a small dip buried  in the raw light curve (see the bottom panel of Figure~\ref{fig:lc5}). We removed instrumental trends from the light curve as described in Section~\ref{sec:lc_ext} and use the  \texttt{batman} Python package to estimate the transit depth and scatter in the transit signal. The transit depth $\delta \approx 0.0011$ (1,100 ppm) estimated from the white-light curve shown in Figure~\ref{fig:detrends} corresponds to a $20.7\sigma$ detection, where $\sigma \sim 5.3 \times 10^{-5}$ is the standard deviation of the residuals. This implies a highly significant transit detection.

Unlike distant exoplanets, the fraction of sunlight blocked by Venus during transit depends on its apparent angular area relative to the Sun, i.e, $\delta = \left ( \theta_{\text{\venus}}\text{/}\theta_\odot \right )^2$, where $\theta_{\text{\venus}} $ and $ \theta_\odot $ are the apparent angular radii of Venus and Sun during the transit. We determine $\theta_{\text{\venus}}\approx 30^{\prime\prime}$ which is consistent with the value reported by \cite{Reale_2015}. 

In Appendix A, we derive a general expression \ref{Eqs:A3} for the transit depth $\delta$ as a function of $R_\star$, $R_p$, $a$, and the observer's distance $d_\star$ which reduces to planet-star area ratio, $R_p^2\text{/}R_\star^2$, when $d_\star \gg a$. If Venus were observed as an exoplanet around a Sun-like star, the expected transit depth would be $\sim$78~ppm\footnote{We use $R_{\text{\venus}} =$ 6131~km \citep{Reale_2015} and $R_\odot =$ 695900~km (from HMI/SDO header, RSUN\_REF).}, whereas, the SDO's measurement of ToV gives $\delta\sim 1,100 $~ppm which is comparable to Neptune-class transit in a distant exoplanetary system. 

\begin{figure}
    \centering
    \includegraphics[width=\columnwidth]{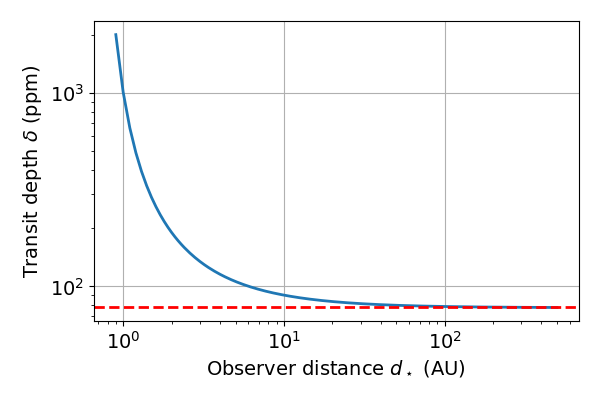} \vspace{-0.5cm}
    \caption{Transit depth of Venus (in ppm) as a function of observer distance. The horizontal red-dashed line represents a 78~ppm mark.}
    \label{fig:ven_dist}
\end{figure}

\begin{figure}
    \centering
    \includegraphics[width=\columnwidth]{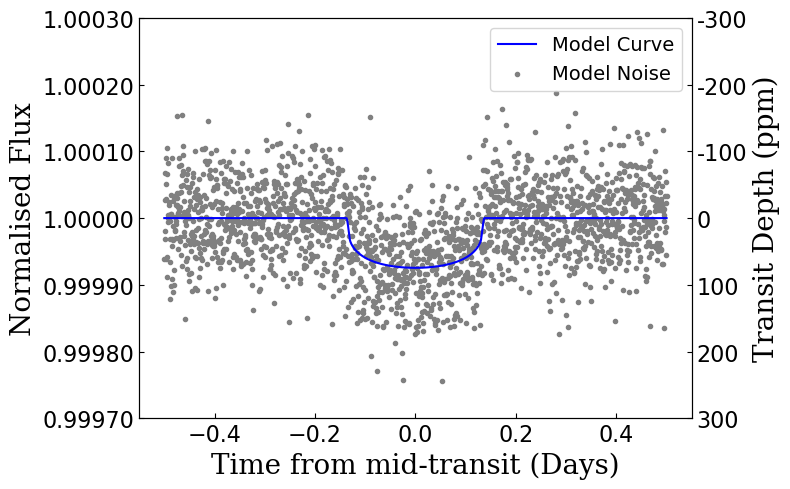}
    \caption{Simulation of ToV as an exoplanetary transit. The model light curve is represented by the blue line and the injected noise by grey dots, assuming a constant signal to noise ratio.}
    \label{fig:ven_from_1000}
\end{figure}

Figure~\ref{fig:ven_dist} shows the transit depth of Venus (in ppm) as a function of observer distance, computed using Equation~\ref{Eqs:A3}. The curve asymptotically approaches $\sim$78 ppm, the exoplanetary limit \citep{Boruki_1984}. This implies that to study solar-system objects as true exoplanetary analogs, one requires a vantage point beyond $\sim$100 AU. Accordingly, any interplanetary probe that ventures into the outer solar system would provide a unique opportunity to observe solar-system transits as genuine exoplanetary analogs. However, a robust detection of the ToV from such a vantage point would remain challenging, since the transit depth declines while photometric noise remains comparable or increases, as demonstrated in the synthetic light curve shown in Figure~\ref{fig:ven_from_1000}.

The Kepler mission was specifically designed to search for Earth-size planets around Sun-like stars using the transit technique \citep{Bor_2010, Koc_2010}. However, the number of such detections fell short of expectations, in part because the mission lifetime was curtailed by reaction wheel failure, and because the photometric variability of most stars exceeded that of the Sun \citep{Gil_2011, Chr_2012}. These limitations underscore why detecting Venus- or Earth-analogs via transits remains challenging.

\subsection{Light curve from AIA UV channel}
We further extend our analysis to the data from the AIA channels in 1700~\AA, 304~\AA, 171~\AA\, and 94~\AA. We find that the transit dip cannot be distinguished from the photometric variations due to the high stellar activity in the 3~EUV wavelengths 304~\AA, 171~\AA\, and 94~\AA~(as previously discussed, see Figure~\ref{fig:lc5}). 

\begin{figure}
  \centering
    \includegraphics[width=\linewidth]{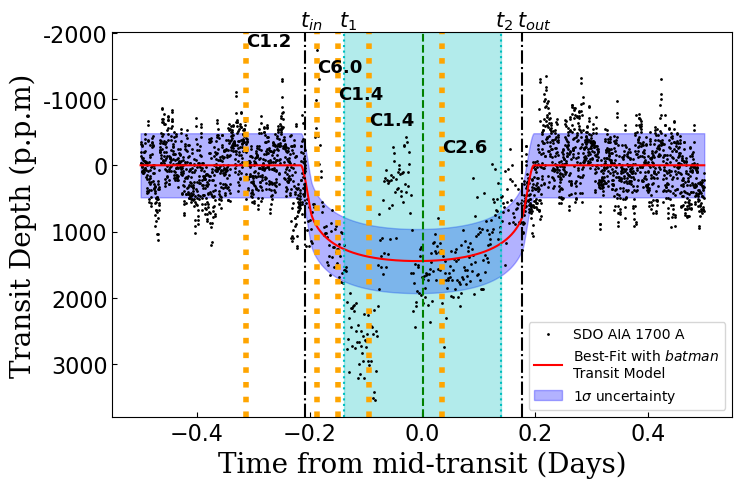}\label{fig:lc1700}
  \vspace{0pt}   
    \includegraphics[width=\linewidth]{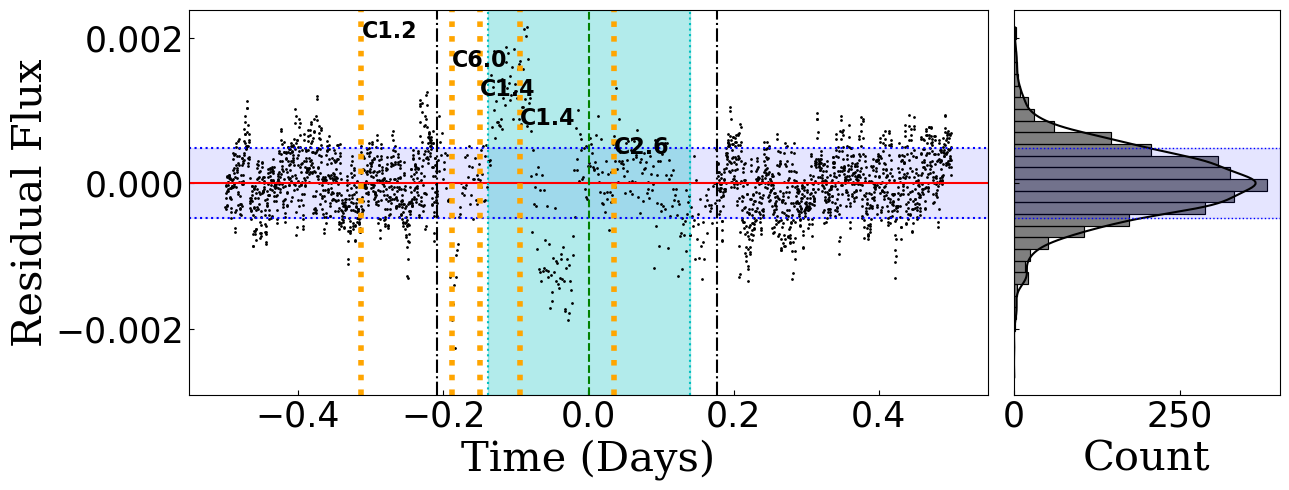}\label{fig:residue1700}
    
  \caption{Transit light curve in the 1700~\AA\ band (top). The red curve shows the best-fit model with the blue shaded region denoting the $1 \sigma = 4.8 \times 10^{-4}$  uncertainty. Black dash-dotted lines mark Venus's entry $t_{\textrm{in}}$ and exit $t_{\textrm{out}}$ from AIA's FoV while the cyan-shaded region between $t_{1}$ and $t_{2}$ indicates the transit in white light. The green dashed line marks the mid-transit time. Bottom panel: residuals of the 1700~\AA\ light curve with 1$\sigma$ bounds shown in blue. Yellow dashed lines indicate flare timings and relative strengths.}
  \label{fig:lc_and_residue_1700}
\end{figure}

In contrast, the light curve obtained from the 1700~\AA\ band after detrending suggests a longer apparent transit dip, as illustrated in Figure~\ref{fig:lc_and_residue_1700}. The plot spans one day, with each data point representing the integrated flux over the AIA full CCD frame after applying Level 1.5 corrections using \texttt{aiapy}. Unlike the light curve constructed from the 6173~\AA\ channel, where flux is integrated only within the visible solar disk, the 1700~\AA\ flux integration was performed over the entire CCD frame to capture Venus occulting the extended UV coronal emission. 

The light curve in Figure~\ref{fig:lc_and_residue_1700} is centered on the mid-transit time, indicated by the green dashed line at approximately 01:29~UTC on 2012-06-06. Since Venus begins to occult the extended UV emission as soon as it enters the SDO/AIA FoV  the UV dip lasts longer than the visible transit, closely tracking Venus entering at $t_\textrm{in}$ and exiting the FoV at $t_\textrm{out}$. The visible transit, lasting about 6 hours 40 minutes, is highlighted in Figure~\ref{fig:lc_and_residue_1700} by the cyan region between $t_1$ and $t_2$, while the period during which Venus remained within the AIA FoV is marked by black dash-dotted lines, extending to about 9 hours 15 minutes which is nearly 40\% longer.

Due to longer path during the approach, Venus required more time to make first contact with the visible disk after entering AIA's FoV than to leave it after the last contact\footnote{$t_1-t_{\textrm{in}}\approx 1.67 \textrm{ hrs}$  and  $t_{\textrm{out}}-t_2\approx 55 \textrm{ mins} $}, i.e., $(t_1-t_{\textrm{in}}) > (t_{\textrm{out}}-t_2)$. Solar flares occurring during the transit are marked by yellow dotted lines, with their respective strengths labeled to the right.

To quantify the UV transit, we modeled the 1700~\AA\ light curve using \texttt{batman}, which indicates a transit depth of $\delta\approx0.0012$, corresponding to a radius of $R_{\text{\venus}} \approx0.035~R_\odot$, or an absolute radius of $\sim$6200~km.\footnote{This agrees with the radius of Venus at 1700~\AA\ UV band \citep{Reale_2015}.} However, unlike in the 6173~\AA\ band, the transit signal at 1700~\AA\ is not very robust ($\delta\approx 2.5 \sigma$) due to high UV variability and sensitivity to flares and plages, which introduce spikes in the flux. \cite{Simones_2019} discuss the spectral content of the UV AIA channels, especially the effect of plages and solar flares, in more detail. Two C-class flares occurred before the transit while Venus was within the AIA FoV, one of which was C6.0, causing a spike in intensity. During the ToV, two more C-class flares contributed to fluctuations in the solar flux.\footnote{The solar flare strengths and timings were derived from NOAA. The strengths were updated according to the new science scale by dividing them by 0.7 according to GOES/NOAA \citep[see][]{Janssens_2025}.}

The longer transit duration observed at UV wavelengths during the Venus transit suggests that similar observations of exoplanet transits could, in principle, be used to probe the extended ultraviolet atmospheres or coronae of their host stars. We use simulation to further explore this possibility in the following section.

\section{Transit simulations across the Sun}\label{sec:tran_sim}
The extended duration of the transit observed in the 1700~\AA\ channel suggests the presence of off-limb coronal features occulted by Venus. To explore this further, we generated an ensemble of synthetic transit light curves by injecting a planet into SDO solar images, scaled to the observed solar angular radius. The primary goal of this simulation is to demonstrate the impact of brightness inhomogeneity across the solar disk during three epochs of the solar cycle -- maximum, intermediate, and minimum -- representing different levels of solar activity.

\begin{figure*}
    \centering
    \includegraphics[width=.8\linewidth]{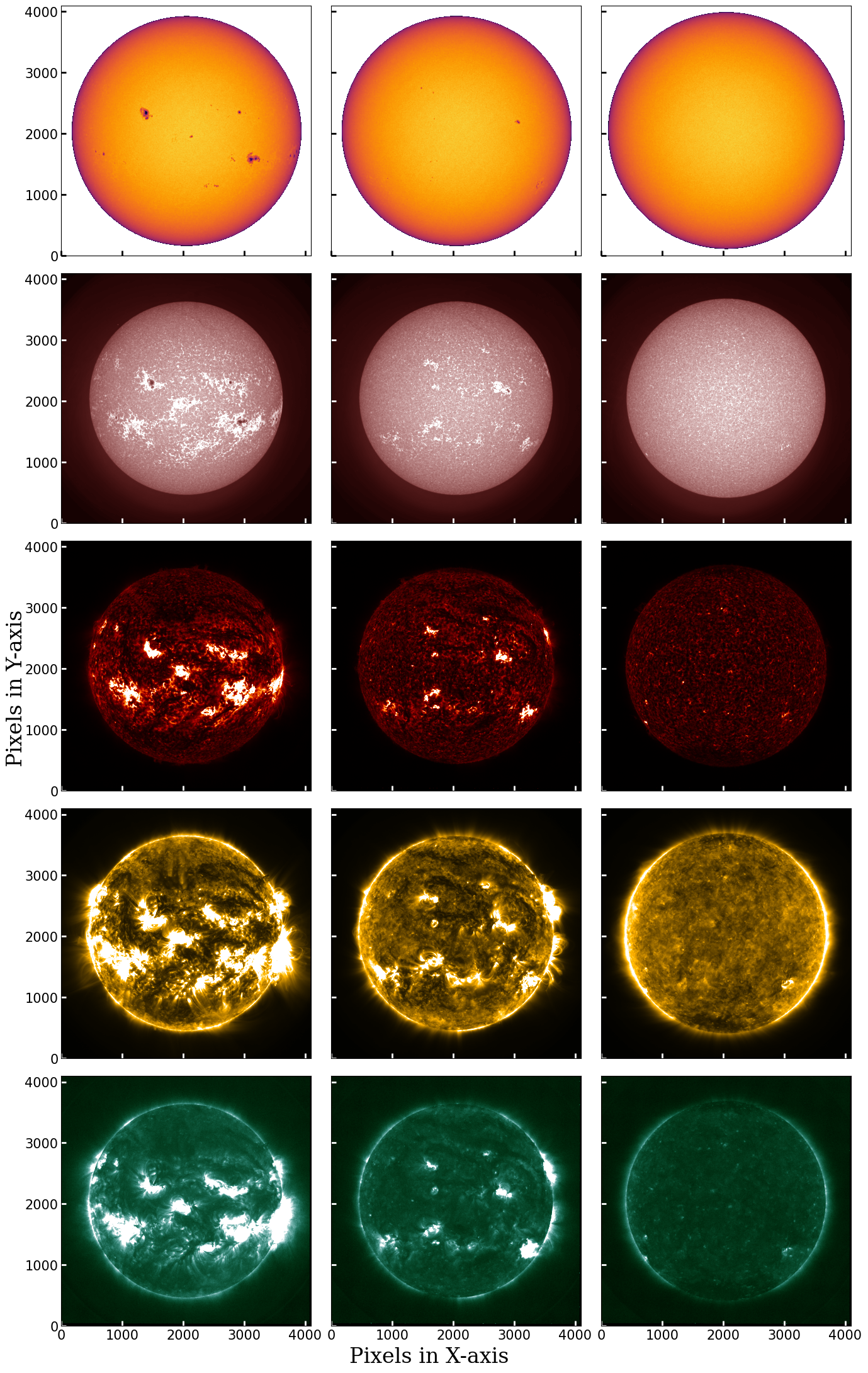}
    \caption{Full-disk SDO solar images chosen for transit simulation. \textit{Left to right}: solar maximum, intermediate, and minimum activity. \textit{Top to bottom:} observations in 6173~\AA, 1700~\AA, 304~\AA, 171~\AA\ and 94~\AA.}
    \label{fig:sun_sim}
\end{figure*}

We selected three representative periods from Solar Cycle~25, which has exhibited higher overall activity than the preceding cycle and has been continuously monitored by SDO. Specifically, full-disk images were randomly chosen in all five wavelengths from three epochs corresponding to the solar maximum (13 August 2024), intermediate phase (19 August 2022), and solar minimum (1 January 2020). Each of these images were corrected to the 1.5 level and the EUV wavelengths were deconvolved with their point spread function with \texttt{aiapy} \citep{Poduval_2013,Barnes2020}. We also check for saturated pixels in each image with the \texttt{NSATPIX} keyword in the AIA headers. We find that only one image has a maximum of 20 pixels, which is negligible \citep{Llama_2015}. 
Figure~\ref{fig:sun_sim} shows the fifteen randomly selected full-disk images, arranged such that columns correspond to the three epochs and rows to the five wavelengths.
Next, for each epoch and wavelength, we simulated transits of three planets corresponding to size of Venus, Neptune, and Jupiter. The radii of Venus, Neptune, and Jupiter were scaled to 0.0087, 0.0356, and 0.1028 times the solar angular radius\footnote{Values taken from the header keyword \texttt{RSUN\_OBS}.}, respectively. For every combination of epoch, wavelength, and planet size, 30 transits were simulated with the planet crossing the solar disk center at random orientations. The corresponding light curves were generated by shifting a dark circular mask\footnote{All pixels within the circle were set to zero.} of the planetary radius along the solar disk and integrating the total image intensity at successive positions. A total of 100 steps spanning solar disk and off-limb distance ($\sim1.3R_{\odot}$)  were used to construct each light curve. 

\begin{figure*}
    \centering
    \includegraphics[width=\linewidth]{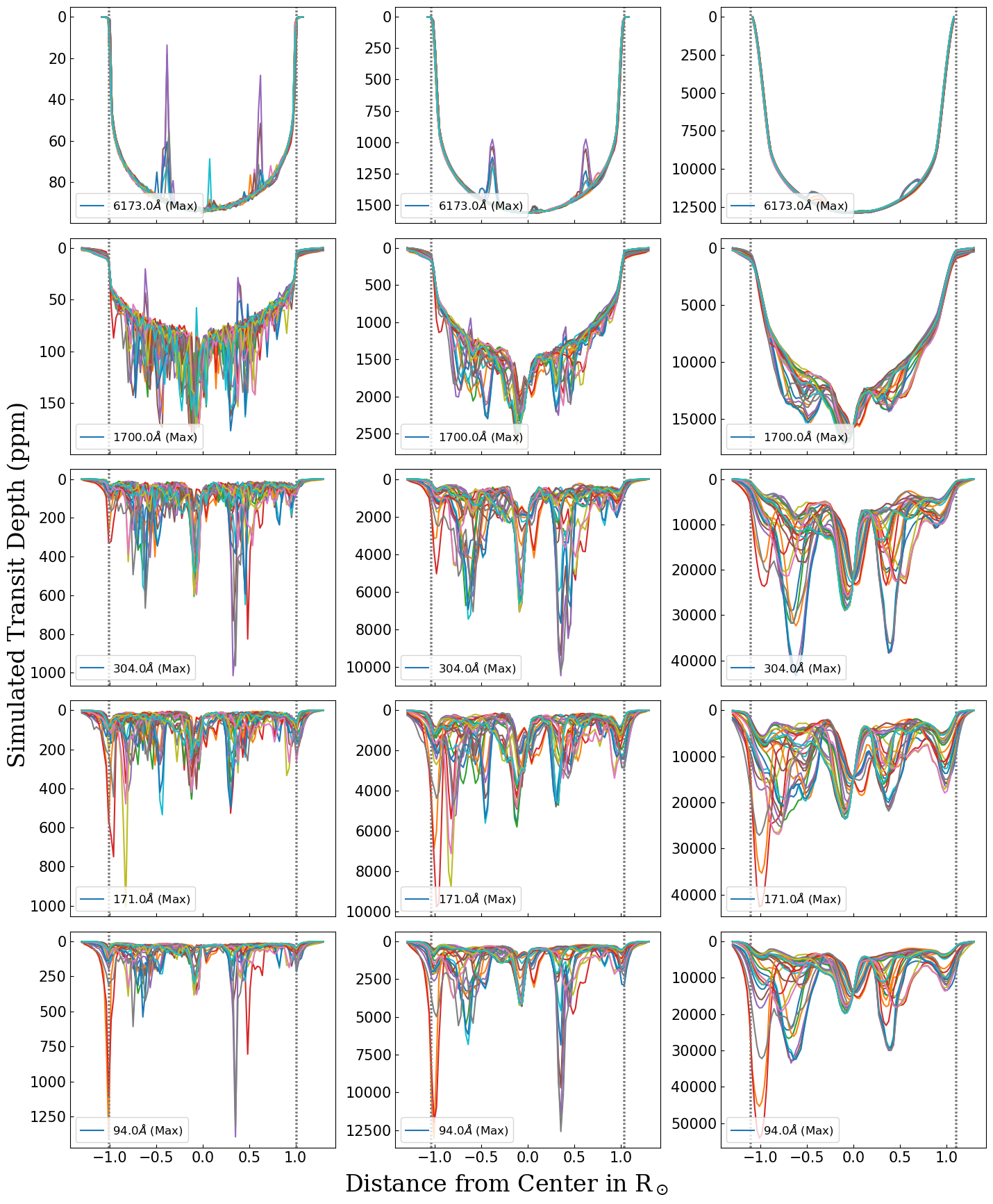}
    \caption{Simulation of transit light curve for Earth/Venus (left column), Neptune (middle column) and Jupiter (right column) size planets in different wavelengths ({\it Top to bottom rows: 6173~\AA, 1700~\AA, 304~\AA, 171~\AA, 94~\AA}) across the Sun during the solar maxima. Each colored curve represents a different angular orientation of the transit path on the solar disk. The vertical dotted lines demarcate the points of ingress and egress of the planet on the visible solar disk. The x-axis represents the distance between the center of the Sun and the planet.}
    \label{fig:max_sim}
\end{figure*}

\begin{figure*}
    \centering
    \includegraphics[width=\linewidth]{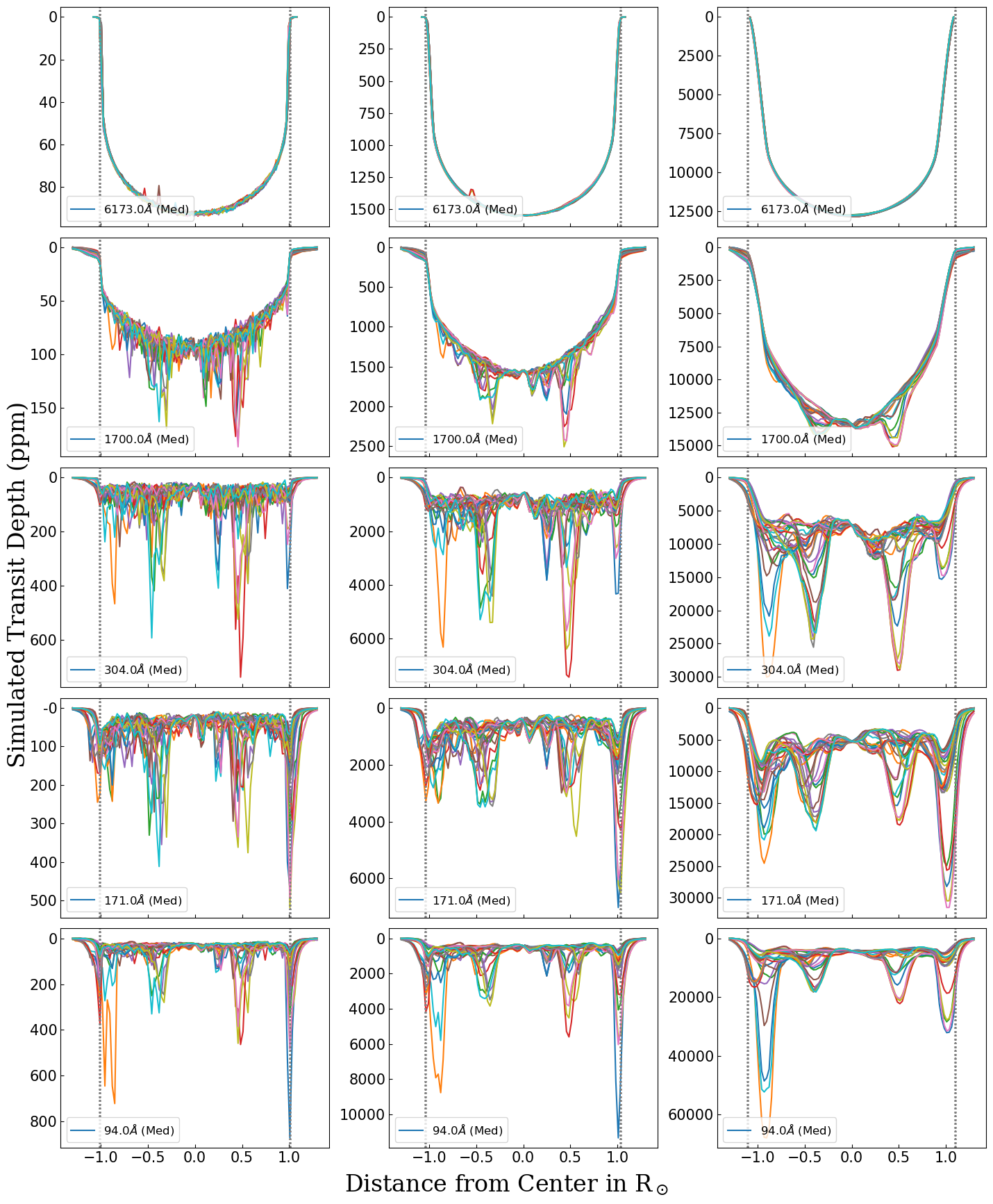}
    \caption{Same as Figure~\ref{fig:max_sim} but during the intermediate stage of the solar cycle.}
    \label{fig:med_sim}
\end{figure*}

\begin{figure*}
    \centering
    \includegraphics[width=\linewidth]{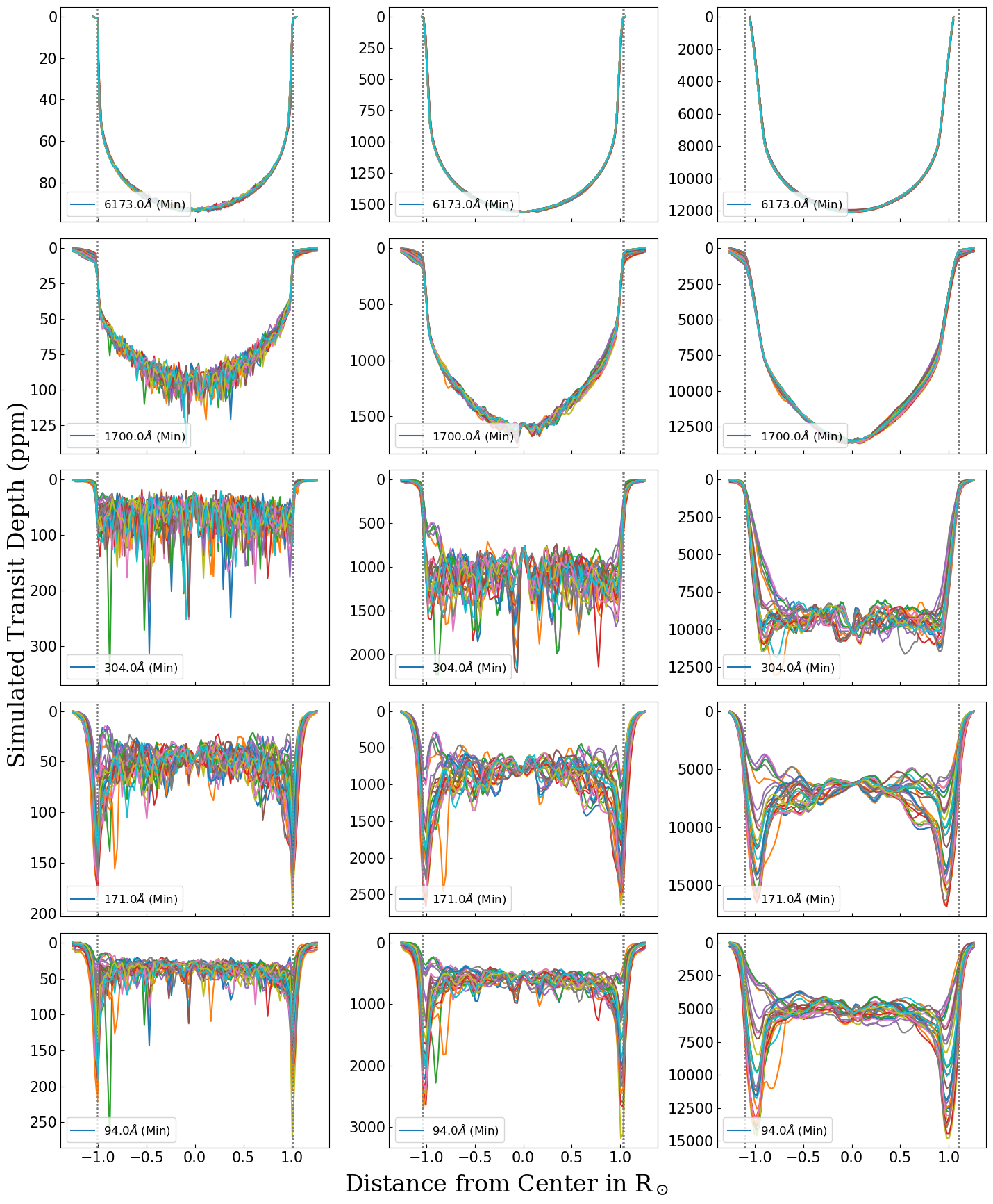}
    \caption{Same as Figure~\ref{fig:max_sim} but during the solar minima, i.e., quiescent phase.}
    \label{fig:min_sim}
\end{figure*}

Results of the synthetic light curves obtained for three activity-cycle epochs—maximum, intermediate, and minimum—are shown in Figures~\ref{fig:max_sim}, \ref{fig:med_sim}, and \ref{fig:min_sim}, respectively. In each figure, the left, middle, and right columns correspond to transits of Venus, Neptune, and Jupiter, respectively, while the rows represent different wavelengths. Because our simulation applies a moving occultor over a static frame, the resulting light curves do not capture slow, time-dependent instrumental systematics or intrinsic stellar brightness variations that would develop over a real observing sequence. Nevertheless, these light curves effectively reveal the brightness inhomogeneities across the solar disk and off-limb regions. Below, we summarize the key points from these simulations that are broadly applicable to the Solar System as well as to exoplanetary transit studies.

\begin{itemize}

    \item \textit{Ingress/egress asymmetry and extended emission:} At UV/EUV wavelengths, transits often begin earlier and end later than their optical counterparts because the occultor intercepts extended off-limb emission. This results in a systematic lengthening of the apparent transit duration—an “early ingress / late egress” signature that reflects the contribution of the extended solar atmosphere rather than the photospheric disk alone. As shown in Figures~\ref{fig:max_sim}, \ref{fig:med_sim}, and \ref{fig:min_sim}, the extended UV/EUV transit duration is evident from its offset relative to the vertical dotted lines which indicate planetary ingress and egress on the visible solar disk. For the 1700~\AA\ we find that these offsets form a band-like structure due to the overlap of several transits, further indicating that the region outside the solar disk is also non-uniform.

    \item \textit{Active-region occultations (negative dips vs. positive bumps):} When the occultor transits over bright active regions at short wavelengths, it produces pronounced  dips in the light curve. Conversely, occultation of dark photospheric spots at optical wavelengths yields localized  bumps. Small-scale fluctuations arising from photospheric granulation are also discernible in the top panel of Figure~\ref{fig:min_sim} and discussed by \cite{Chiavassa_2017}. Toward shorter UV/EUV wavelengths, the light curves become increasingly jagged, further demonstrating the pronounced surface-brightness inhomogeneity dominated by plages and faculae during the Sun's active and intermediate phases. These results are in agreement with the existing literature as well which suggest similar occurrences \citep{France_2023,Valio_2024,Araujo_2025}.

    \item \textit{ Planet-size and SNR dependence:} The magnitude of activity-induced deviations in the light curve scales with occultor size. Smaller planet (e.g. Venus) sample local brightness inhomogeneities more strongly (larger fractional scatter), while larger occultors (e.g. Neptune or Jupiter) average over more structure and show smoother, but still systematically biased, light curves. Consequently, retrieved planet-to-star radius ratios $(R_p\text{/}R_\star)$ at short wavelengths can be biased by tens of percent depending on whether bright regions are occulted \citep{Llama_2015,King_2024}.
    
    \item \textit{ Optical versus short-wavelength morphology and limb-brightening.} The simulated 6173~\AA\ light curves exhibit the expected limb-darkened morphology and comparatively small scatter, consistent with a nearly homogeneous photospheric intensity distribution. In contrast, UV and EUV light curves deviate markedly from classical limb-darkening, displaying limb-brightened profiles and wavelength-dependent departures caused by spatially localized bright regions (plages, active regions) and off-limb coronal emission. This limb-brightening becomes particularly prominent during the Sun's quiet phase, producing a characteristic `W-shaped' transit signature (see Figure~\ref{fig:min_sim}). The nearly featureless solar disk at minimum activity shows a diffuse glow along the limb, which is more pronounced at lower latitudes that at polar regions as evident in the bottom panel of Figure~\ref{fig:sun_sim}.
    
    \item \textit{ Activity level and transit detectability:} Simulations across different activity epochs—maximum, intermediate, and minimum, show that both the amplitude and frequency of jagged features in the UV/EUV light curves increase with activity level. During solar maximum, the transit profiles become highly erratic, whereas during quiescent phase they approach the idealized limb-darkened or limb-brightened forms. Likewise, for highly active stars, the strong spatial and temporal brightness variations in the UV/EUV bands would make it difficult to detect transits even for Jupiter-sized planets. In contrast, during the low-activity quiescent phase, it may still be possible to discern transits of planets larger than Neptune, especially in EUV where the limb-crossing signal is conspicuous.  Temporal variability during actual stellar observations could further hinder the detection of smaller (e.g., Venus-sized) planets at 1700~\AA, while Neptune- and Jupiter-sized planets remain within reach.
    
\end{itemize}

\section{Conclusion}\label{sec:con}


We have analyzed the 2012 transit of Venus using full-disk SDO/HMI and AIA observations in five wavelength channels -6173~\AA\  (continuum), 1700~\AA\ (UV), and three EUV bands at 304, 171, and 94~\AA. The visible (6173~\AA) light curve shows a well-defined transit with a depth of $\delta\approx 0.0011$ ($\approx 1100$~ppm), corresponding to an apparent angular radius of Venus $\sim30\arcsec$ ($R_{\text{\venus}} \text{/} R_\odot \approx 0.033$), and a duration consistent with the predicted ephemerides, 6 hours 40 minutes. In the 1700~\AA\ band, however, the transit appears 40\% longer (by $\sim2$~hours 35 mins), as Venus begins occulting UV-emission from solar coronal before reaching the optical limb. This provides direct observational evidence of the early-ingress/late-egress effect predicted in previous studies \citep{Llama_2015}. In contrast, the EUV light curves are dominated by strong solar variability and flaring activity, preventing a clear transit detection. This is unsurprising, as the 2012 transit of Venus occurred near the peak of Solar Cycle 24, when the Sun was in a relatively active phase.

When viewed as an exoplanet analog, the observed depth is much larger than the $\sim78$~ppm expected for a distant Venus-like planet, implying that a vantage point well beyond 100~AU is needed for solar system transits to mimic the true exoplanetary analogs. 

We also carried out synthetic transit simulations over quiet, intermediate, and active-Sun conditions. These numerical experiments  show that UV and EUV transits differ significantly from their optical counterparts due to limb brightening and the spatial inhomogeneity of the solar surface and corona. The resulting light curves are highly structured, with bright plages, active regions, and off-limb emission introducing strong variability, particularly during solar maximum. Small planets, such as Venus, are more sensitive to these localized variations, while larger planets produce smoother yet still biased light curves. The detectability and morphology of UV/EUV transits are thus strongly activity-dependent, appearing more stable and symmetric during quiet-Sun epochs, but becoming irregular and often undetectable during active phases.

In summary, our multiwavelength analysis demonstrates that UV transit photometry offers a unique diagnostic to probe the extended structure of stellar atmospheres. The longer duration of the 1700~\AA\ Venus transit confirms that planets can occult chromospheric and coronal layers beyond the optical photosphere, while EUV regimes remain noise-dominated for reliable transit detection on Sun-like stars. Comparing transit depth and duration across visible and UV wavelengths can thus reveal the presence and extent of coronal emission, providing valuable insight into stellar and planetary atmospheres.

These results can have important implications for exoplanet transit science. Future space missions should include dedicated UV photometric monitoring, through instruments like HST or upcoming missions such as ULTRASAT, to exploit this capability \citep{Shv_2024, Man_2025}. In particular, the unambiguous `W-shaped' limb-crossing event during the quiescent phase of the star can serve as a strong telltale signature of giant planet detection at EUV wavelengths. However, further modeling of time-dependent stellar variability and extending such analyses to other stellar types will be essential. High-precision UV light curves of transiting planets, especially those possessing extended or escaping atmospheres, would significantly advance the study of exoplanetary atmospheres.

\section*{Acknowledgement}
The courtesy of NASA/SDO and the AIA and HMI science teams. NOAA Office of Satellite and Product Operations (1994): NOAA Geostationary Operational Environmental Satellite (GOES) I-M and N-P Series Imager Data, June 4-8, 2012. NOAA National Centers for Environmental Information. \href{doi:10.25921/Z9JQ-K976}{doi:10.25921/Z9JQ-K976}. August 1 2025.

\section*{Data Availability}
The full-disk solar images analyzed in this study were obtained from the SDO data archive maintained by the Joint Science Operations Center at Stanford University (\href{http://jsoc.stanford.edu/}{http://jsoc.stanford.edu/}). Solar flare event data were retrieved from the Space Weather Prediction Center of the National Oceanic and Atmospheric Administration (\href{https://cesar.kso.ac.at/database/swpc\_flares\_query.php}{https://cesar.kso.ac.at/database/swpc\_flares\_query.php}). The  disk-integrated light curves generated for this work are available from the corresponding author upon reasonable request.

\bibliographystyle{mnras}
\bibliography{biblio}

\appendix 

\section{Transit depth in exoplanets versus solar system objects}

For completeness, we note here an important distinction between the manifestation of transit depth in exoplanetary systems and in solar system transits. For transit observations within solar system, the sun and planet are close enough that their apparent disks are well resolved when viewed from Earth. In general, the fractional loss of stellar flux $\delta$ during a transit is determined by the ratio of the projected angular areas of the planet and the star  as seen by the observer. 

\begin{figure}
    \centering
    \includegraphics[width=0.8\columnwidth]{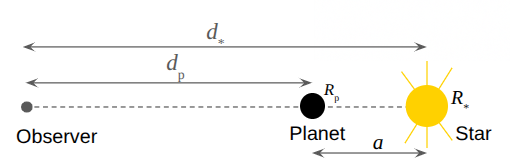}
    \caption{Transit geometry for object at closer distance to observer.}
    \label{fig:apx_fig1}
\end{figure}

Let $R_p$ and $R_\star$ denote the planetary and stellar radii, respectively, and let $d_p$ and $d_\star$ be their distances from the observer (see Figure~\ref{fig:apx_fig1}). The apparent angular radii are then given by:
\begin{equation}
    \theta_p = \frac{R_p}{d_p}, \;\;\;\; \theta_\star = \frac{R_\star}{d_\star}
\end{equation}

The corresponding transit depth is:

\begin{equation}
    \delta= \left ( \frac{\theta_p}{\theta_\star} \right)^2 = \left ( \frac{R_p}{R_\star} \right )^2 \left(   \frac{d_\star}{d_p}\right )^2 
\end{equation}
If the semi-major axis of the planetary orbit is $a$, then during transit the line-of-sight distance to the planet is: \(d_p = d_\star-a\).
Therefore, the transit depth becomes: 
\begin{equation}
    \delta= \left ( \frac{R_p}{R_\star} \right )^2 \left( 1- \frac{a}{d_\star}\right )^{-2} 
    \label{Eqs:A3}
\end{equation}
Equation~\ref{Eqs:A3} provides a general expression for estimating the transit depth of any solar system objects observed from a another location within the solar system.
For distant planetary systems, $d_\star>> a$, therefore, 
\begin{equation}
    \delta \approx \left( \frac{R_p}{R_\star} \right)^2 
\end{equation}
which is routinely used in exoplanet studies. Thus, both solar system and exoplanet transits are described by the same general framework, with the difference arising only from whether the observer is close enough to resolve the planet and the stellar disks.

\bsp	
\label{lastpage}
\end{document}